\magnification 1200
\centerline {\bf Local Thermodynamic Equilibrium at  Three Levels}
\vskip 1cm
\centerline {by Geoffrey L. Sewell}
\vskip 0.5cm
\centerline {Department of Physics, Queen Mary University of London}
\vskip 0.3cm
\centerline {(e-mail: g.l.sewell@qmul.ac.uk)}
\vskip 0.5cm
\centerline {Mile End Road, London E1 4NS, UK}
\vskip 1cm
\centerline {\bf Abstract}
\vskip 0.3cm
We present coordinated formulations of local thermodynamical equilibrium 
conditions at three levels, namely the macroscopic one of classical thermodynamics, 
the mesoscopic one of hydrodynamical fluctuations and the microscopic one of 
quantum statistical mechanics. These conditions are all expressed in terms of the 
hydrodynamical variables of the macroscopic picture, and the quantum statistical ones 
are shown to imply a local version of the zeroth law. 
\vfill\eject
\centerline {\bf 1. Introduction}
\vskip 0.3cm 
The concept of local thermodynamic equilibrium (LTE) is basic, at the macroscopic 
level, to nonequilibrium thermodynamics [DM, LL]; while at the microscopic level, it 
features in certain statistical mechanical models, such as the quantal ones of  [Da, 
NY] and the classical ones of [KMP, DIPP, Pr]. However, there does not appear to be 
a general, coherent formulation of this concept that coordinates the pictures of LTE at 
the macroscopic level of classical nonequilibrium thermodynamics, the mesoscopic 
one of hydrodynamical fluctuations and the microscopic one of quantum statistical 
mechanics. The object of this note is to present such a formulation and to show that its 
quantum statistical component leads to a local version of the zeroth law. This 
objective marks part of our programme [Se1-3] of building bridges between the 
quantum microscopic and classical macroscopic pictures of matter, rather than an 
attempted derivation of the latter from the former.
\vskip 0.2cm
We base our treatment on the generic model of a nonrelativistic many-particle 
quantum system, ${\Sigma}$, with translationally  invariant interactions, which 
occupies an open, connected region of a $d$-dimensional Euclidean space, $X$, and 
is coupled at its boundary to an array, ${\cal R}$, of reservoirs. We assume, in a 
standard way [Ca], that its equilibrium thermodynamics is based on the form of an 
entropy function, $S$, of a set $Q \ \bigl(=(Q_{1},. \ .,Q_{n})\bigr)$ of extensive 
conserved variables, of which $Q_{1}$ is the energy. Assuming that $S(Q)$ is also 
extensive, its density is a volume independent function, $s$, of $q$, the density of 
$Q$. This latter density represents the state of ${\Sigma}$ at the thermodynamic 
level. 
\vskip 0.2cm
For the nonequilibrium situation, we assume that the classical continuum mechanics 
of the model is given by an autonomous evolution of the position and time dependent 
density, $q(x,t)$, of $Q$, which takes the form\footnote*{This assumption is 
manifestly satisfied by diffusion and heat conduction processes, even nonlinear ones, 
as well as by Navier-Stokes hydrodynamics.}
$${{\partial}\over {\partial}t}q(x,t)={\Phi}(q:x,t),\eqno(1.1)$$
where ${\Phi}$ is a functional of $q$. We assume, for simplicity, that this equation is 
invariant under space-time scale trransformations of the form 
$x{\rightarrow}{\lambda}x, \ t{\rightarrow}{\lambda}^{c}t$, where ${\lambda}$ is 
a variable positive parameter and $c$ is a positive constant. This assumption is 
satisfied in the cases of nonlinear diffusion and inviscid Eulerian hydrodynamics. In 
the former case,  ${\Phi}(q)$ takes the form ${\nabla}.(K(q){\nabla}q)$, with $K$ a 
positive $n$-by-$n$ matrix valued function of $q$  and $c=2$: in the latter case, it is 
easily seen from the Euler equations that $c=1$. On the other hand, the assumption is 
not satisfied by Navier-Stokes equations for viscous flow: this case requires a 
modified treatment [Se3] that we shall briefly discuss in Section 5. 
 \vskip 0.2cm 
In order to treat ${\Sigma}$ on different levels of macroscopicality, we need to 
specify the relationships between the length  and time scales for those levels. Our 
choice is to take the unit of length for the macroscopic and mesoscopic pictures to be 
$L$ times that of the microscopic one, where $L= N^{1/d}, \ N$ being the number of 
particles in ${\Sigma}$.Thus, denoting by ${\Omega}$ the region of $X$ occupied 
by ${\Sigma}$ on the macroscopic scale, this same region is 
${\Omega}_{L}=L{\Omega} \ 
\bigl(={\lbrace}Lx{\vert}x{\in}{\Omega}{\rbrace}\bigr)$ on the microscopic one. 
We take ${\Omega}$ to be $L$-independent and  choose the macroscopic units so 
that its volume is unity: the volume of ${\Omega}_{L}$ is therefore $L^{d}=N$.
\vskip 0.2cm
In view of the assumed scale invariance of the macroscopic dynamics of the model, 
specified following Eq. (1.1), the choice of $L$ for the ratio of the macroscopic unit 
of length to the microscopic one demands that the corresponding ratio for the units of 
time is $L^{c}$. Hence the time scales for the microscopic and macroscopic pictures 
are infinitely separated in the hydrodynamic limit $L{\rightarrow}{\infty}$. The 
system therefore supports dynamics over two quite different time scales, namely a 
macroscopic one for which we continue to denote the time variable by $t$ and a 
microscopic one for which we denote that variable by ${\tau}$. Indeed, the 
macroscopic time may be considered to be \lq frozen\rq \ at a value $t$ while the 
system undergoes its quantum evolution on the microscopic time scale. Hence, it is 
natural to anticipate that a submacroscopic\footnote*{Here a submacroscopic region 
is one that is extremely small in macroscopic terms but still large enough to contain 
an enormous number of particles.} region of the system concentrated around a point 
$x$ of ${\Omega}$ is driven by its microscopic kinetics to a state of equilibrium 
corresponding to the value of $q(x,t)$. This is conceived, heuristically, to be the 
essential mechanism whereby local equilibrium is generated at all levels.
\vskip 0.2cm
In addition to $L$, other key parameters of the model are the Planck and Boltzmann 
constants, ${\hbar}$ and $k$. We assume that, in the units of the macroscopic picture, 
these two quantities  are extremely small\footnote{**}{For example, in SI units, 
${\hbar}$ and $k$ are of the order of $10^{-34}$ and $10^{-23}$ respectively.}. 
Accordingly, we employ limits where ${\hbar}$ and $ k$, as well as $L^{-1}$, tend 
to zero in  certain parts of our treatment. Specifically, the limit 
${\hbar}{\rightarrow}0$ is implicit in the assumption of macroscopic classicality that 
underlies the phenomenological picture and which we extend to the mesoscopic one 
of hydrodynamical fluctuations\footnote{***}{This supplementary assumption is 
consistent with the known result [GVV] that the fluctuation fields satisfy 
commutation relations, which are applicable to both classical and bosonic fields}; the 
limit $L{\rightarrow}{\infty}$ is standard for the connection between microscopic 
and  hydrodynamic properties of matter; and the limit $k{\rightarrow}0$, which we 
term the {\it Boltzmann limit}, arises in our formulation of hydrodynamic 
fluctuations, which for equilibrium states are governed by Einstein\rq s relation, 
$P={\rm const'} \ {\rm exp}(S/k)$, between their probability distribution, $P$, and 
the entropy $S$.
\vskip 0.2cm 
We begin our treatment in Section 2 by formulating largely standard macroscopic, 
mesoscopic and microscopic pictures of {\it global} thermal equilibrium of the model 
in terms that facilitate a subsequent passage to corresponding pictures of LTE. For the 
macroscopic description, in Section 2.1, we express the thermodynamics of the model 
in terms of the entropy density $s(q)$ and its Legendre 
transform\footnote{****}{This is just the ratio of the pressure to the temperature.}, 
${\pi}({\theta})$, where ${\theta}$, the thermodynamic conjugate of $q$,  is just the 
first derivative of the $s(q)$.  We term it the {\it control variable}. For the mesoscopic 
picture, in Section 2.2, we employ Einstein\rq s formula, in the Boltzmann limit, to 
express the statistical properties of the hydrodynamical fluctuations in terms of the 
Hessian of ${\pi}({\theta})$. For the microscopic picture, in Section 2.3, we 
formulate the model in operator algebraic terms, characterising its equilibrium states 
by a global thermodynamical stability (GTS) condition that corresponds precisely to 
that of the macroscopic description. We then employ the quantum statistical model to 
obtain a rather simple condition for the {\it thermodynamical completeness} of the 
variables $Q$ of the macroscopic picture. Furthermore we recall that the GTS 
condition implies the dynamical one of Kubo-Martin-Schwinger (KMS), subject to 
certain technical assumptions; and consequently [KFGV] that the model conforms to 
the zeroth law, in that it drives finite systems to which it is locally coupled to 
equilibrium at its own temperature. 
\vskip 0.2cm
In  Section 3 we pass from global to local equilibrium conditions in the following 
way. We start, in Section 3.1,  by recalling that the equation of motion (1.1) involves 
the assumption that, at a local level, the system supports a thermodynamics in which 
the entropy density at position $x$ and time $t$ is just the equilibrium entropy density 
function, $s$, of $q(x,t)$,  the local  density of $Q$. This is the local equilibrium 
hypothesis at the macroscopic level and $q(x,t)$ is now the solution of Eq. (1.1), 
subject to the boundary conditions imposed by the reservoirs ${\cal R}$. The 
thermodynamical conjugates ${\theta}(x,t)$ of $q(x,t)$ and ${\pi}$ of $s$ are then 
obtained by Legendre transformation, just as for global equilibrium. In Section 3.2, 
we recall that the equilibrium condition for the hydrodynamical fluctuation process, 
formulated in Section 2. reduces locally to a very simple form, expressed in terms of 
the Hessian of ${\pi}({\theta})$. Accordingly, we assume that, for the nonequilibrium 
situation, the local equilibrium condition takes the same form, with the constant 
${\theta}$ replaced by ${\theta}(x,t)$. In Section 3.3, we consider the quantum 
statistical picture of the system in a region that, in the macroscopic units, is an open 
neighbourhood ${\cal N}({\epsilon},x)$ of $x ({\in}{\Omega})$, that shrinks to the 
point $x$ as ${\epsilon}{\rightarrow}0$. Since this region is ${\cal 
N}_{L}({\epsilon},x):=L{\cal N}({\epsilon},x)$, when viewed on the microscopic 
scale, it becomes infinitely large there, covering the whole of $X$, as 
$L{\rightarrow}{\infty}$. Hence, by employing limits in which first 
$L{\rightarrow}{\infty}$ and then ${\epsilon}{\rightarrow}0$, we achieve a 
description in which the neighbourhood corresponds to a hydrodynamic point in being 
macroscopically small and microscopically large. In this limiting situation, the 
quantum model of ${\Sigma}$, as restricted to this region, reduces to that of an 
infinite system, and the local equilibrium condition is simply that of GTS, and hence 
of KMS, corresponding to the value ${\theta}(x,t)$ of the control variable. Thus we 
arrive at the picture wherein the local equilibrium conditions at all levels are 
determined by the macroscopic one, in that their precise forms are just those of global 
equilibrium, but with the constant valued ${\theta}$ replaced by the position and time 
dependent ${\theta}(x,t)$. 
\vskip 0.2cm
In Section 4, we show, by an adaptation of the argument of [KFGV], that the LTE 
condition at the quantum statistical level implies a local version of the zeroth law in 
that, in the limit $L{\rightarrow}{\infty}$,  its coupling to a finite system 
${\Gamma}$ in a neighbourhood of $x$ drives the latter to equilibrium at the local 
temperature of ${\Sigma}$. 
\vskip 0.2cm
We conclude in Section 5 with a summary of our results and a sketch of how they 
may be extended to cases such as that of Navier-Stokes fluids, where the macroscopic 
evolution is not scale invariant.
\vskip 0.5cm 
\centerline {\bf 2. Equilibrium Thermodynamics}
\vskip 0.3cm
We formulate the equilibrium thermodynamics of ${\Sigma}$ here on macroscopic, 
mesoscopic and microscopic levels.
\vskip 0.5cm 
\centerline {\bf 2.1. The Macroscopic Picture} 
\vskip 0.3cm
In a standard formulation [Ca], the thermodynamics of ${\Sigma}$ may be expressed 
in terms of a set of extensive conserved variables $Q=(Q_{1},. \ .,Q_{n})$ and an 
entropy function, $S$, of these: in particular, $Q_{1}$ is the energy of the system. 
We assume that the range of  $Q$ is a convex subset of ${\bf R}^{n}$ and we note 
that the demand of thermodynamic stability ensures that the function $S$ is concave. 
Further, its value, $S( Q)$, like that of $Q$, is extensive. We shall provide a quantum 
statistically based thermodynamic completeness condition for $Q$ in Section 3.3.
\vskip 0.2cm
The fundamental thermodynamic formula for the change in $S$ incurred in quasi-
static transitions between equilibrium states is 
$$TdS=dE+{\Sigma}_{j=2}^{n}f_{j}dQ_{j}+pdV,\eqno(2.1.1)$$
where $T$ is the temperature, $p$ the pressure, $V$ the volume of ${\Omega}$ and 
$f_{j}dQ_{j}$ is the work done by the system in effecting an infinitesimal change in 
$Q_{j}$. In view of the extensivity of $Q$ and $S(Q)$, these may be expressed as 
$qV$ and $s(q)V$, respectively, where the function $s$ is $V$-independent and, like 
$S$, is concave. Here $q$ is an $n$-tuple $(q_{1},. \ .,q_{n})$ and represents the 
thermodynamic state of ${\Sigma}$. We denote its range by ${\cal Q}$, and we shall 
sometimes denote $q_{1}$, the energy density, by $e$. 
\vskip 0.2cm
Since $Q=qV$ and $S(Q)=s(q)V$, Eq. (2.1.1) is equivalent to the formula
$$\bigl[Tds-de-{\sum}_{j=2}^{n}f_{j}dq_{j}\bigr]V+\bigr[Ts-e-
{\sum}_{j=2}^{n}f_{j}q_{j}-p\bigr]dV=0,$$
from which it follows that
$$Tds=de+{\sum}_{j=2}^{n}f_{j}dq_{j}\eqno(2.1.2)$$
and
$$p=Ts-e-{\sum}_{j=2}^{n}f_{j}q_{j}.\eqno(2.1.3)$$
Note that these formulae and their consequences pertain to the thermodynamic limit 
of the corresponding statistical mechanical ones, since the extensivity relations 
$Q=qV$ and $S(Q)=s(q)V$ represent approximations wherein surface effects are 
neglected.
\vskip 0.2cm
The thermodynamic conjugate of $q$ is the control variable, ${\theta}$, defined by 
the formula 
$${\theta}=({\theta}_{1},. \ .,{\theta}_{n})=s^{\prime}(q):=
({\partial}s(q)/{\partial}q_{1}, . \ .,{\partial}s(q)/{\partial}q_{n}\bigr).\eqno(2.1.4)$$ 
Hence, by Eqs. (2.1.2) and (2.1.4),
$${\theta}_{1}=T^{-1} \ {\rm and} \ {\theta}_{j}=T^{-1}f_{j} \ {\rm for} \ j=2,. \ 
.,n.\eqno(2.1.5)$$
We denote the range of ${\theta}$ by ${\Theta}$, which we term the {\it control 
space} and assume to be convex.  
 \vskip 0.2cm
The thermodynamic potential conjugate to $s$ is defined to be the function, ${\pi}$, 
on ${\Theta}$ given by the formula
$${\pi}({\theta})={\rm sup}_{q^{\prime}{\in}{\cal Q}}\bigl(s(q^{\prime})-
{\theta}.q^{\prime}\bigr),\eqno(2.1.6)$$
where the dot denotes the ${\bf R}^{n}$ inner product. Since $s$ is concave, it 
follows from the definition of ${\Theta}$ that the supremum of this formula is 
attained by a value, $q$, of $q^{\prime}$ related to ${\theta}$ by Eq. (2.1.5), i.e.
$${\pi}({\theta})=s(q)-{\theta}.q.\eqno(2.1.7)$$
We shall presently discuss conditions under which this $q$ is unique. In any case, it 
follows from Eqs. (2.1.3), (2.1.5) and (2.1.7) that
$${\pi}({\theta})=p/T.\eqno(2.1.8)$$
In view of this formula, we term ${\pi}$ the {\it reduced pressure}. It follows from 
Eq. (2.1.6) that ${\pi}$ is convex. Hence, in a standard terminology for convex 
functions, a tangent to the graph of ${\pi}$ at a point $P$ is just a line through $P$ 
that lies below that graph. Thus, the set, ${\cal T}_{\theta}$, of tangents at the point  
$\bigl({\theta},{\pi}({\theta})\bigr)$ consists of the lines through that point whose 
slopes, $r$, satisfy the inequality
$${\pi}({\theta}^{\prime})-{\pi}({\theta}){\geq}r.({\theta}^{\prime}-{\theta}) \ 
{\forall} \ {\theta}^{\prime}{\in}{\Theta}.\eqno(2.1.9)$$ 
It follows immediately from this formula that ${\cal T}_{\theta}$ is convex. We 
denote the set of its extremal elements by ${\cal E}({\cal T}_{\theta})$. Further, by 
Eqs. (2.1.6) and (2.1.7),
$${\pi}({\theta}^{\prime})-{\pi}({\theta}){\geq}-q.({\theta}^{\prime}-{\theta}) \ 
{\forall} \ {\theta}^{\prime}{\in}{\Theta},\eqno(2.1.10)$$ 
which signifies, by Eq. (2.1.9), that $-q{\in}{\cal T}_{\theta}$. Thus, the equilibrium 
states, $q$, corresponding to the control parameter ${\theta}$ are elements of 
$-{\cal T}_{\theta}:={\lbrace}-r{\vert}r{\in}{\cal T}_{\theta}{\rbrace}$. Conversely 
[Se1, Prop. 6.3.1], assuming that this set is contained in ${\cal Q}$, it comprises all 
the equilibrium states, as defined by Eq. (2.1.7). Hence we identify the equilibrium 
states with the elements of $-{\cal T}_{\theta}$. In particular, we 
assume\footnote*{This assumption is supported in Section 2.3 on the quantum 
statistical grounds that the extremals are the only equilibrium states for which $q$ is 
sharply defined (cf. [Ru]): for the others this is a random variable with non-zero 
dispersion.} that the pure phases are represented by the extremals, $-{\cal E}({\cal 
T}_{\theta})$. Thus, in the case where ${\pi}$ is differentiable at the value 
${\theta}$ of the control parameter, ${\cal T}_{\theta}$ consists of the single 
element, $-q$, corresponding to a pure phase and defined by the formula
$${\pi}^{\prime}({\theta})=-q.\eqno(2.1.11)$$
It follows from this equation and Eq. (2.1.4) that the correspondence between $q$ and 
${\theta}$ is one-to-one and therefore that, in the pure phase regime, we may 
equivalently represent the state of ${\Sigma}$ by either of these variables. 
Moreover, assuming that, in this regime, both $s$ and ${\pi}$ are twice 
differentiable, it follows from Eqs. (2.1.4) and (2.1.11) that the Hessians, 
$s^{{\prime}{\prime}}(q) \    
\bigl(:=[{\partial}^{2}s(q)/{\partial}q_{j}{\partial}q_{k}]\bigr)$ 
and ${\pi}^{{\prime}{\prime}}({\theta}) \ 
\bigl(:=[{\partial}^{2}{\pi}({\theta})/{\partial}{\theta}_{j}{\partial}{\theta}_{k}]
\bigr)$ , are related by the equation
$${\pi}^{{\prime}{\prime}}({\theta})s^{{\prime}{\prime}}(q)=
s^{{\prime}{\prime}}({\overline q}){\pi}^{{\prime}{\prime}}({\theta})=-I,$$
i.e. 
$${\pi}^{{\prime}{\prime}}({\theta})=
-[s^{{\prime}{\prime}}(q)]^{-1}.\eqno(2.1.12)$$
\vskip 0.5cm
\centerline {\bf 2.2. The Mesoscopic Picture: Equilibrium Fluctuations}  
\vskip 0.3cm
Here we employ the same spatial scale as for the macroscopic picture, though now the 
density $q$ of $Q$ is a function of position. The statistics of the equilibrium 
fluctuations  are governed by the canonical modification of Einstein\rq s formula for 
the situation where the control variable ${\theta}$ is maintained by the reservoirs at 
the constant value ${\overline {\theta}}$. Assuming that the system is in a pure 
phase, the corresponding value, ${\overline q}$, of $q$ is $-
{\pi}^{\prime}({\overline {\theta}})$, by Eq. (2.1.11). The probability distribution, 
$P$, for the macroscopic variables is then given formally by the equation
$$P={\rm const.}{\rm exp}\bigl([S-{\overline {\theta}}.Q]/k\bigr),$$
i.e.
$$P={\rm const.}{\rm exp}\bigl(k^{-1}\int_{\Omega}dx[s(q(x))-{\overline {\theta}}.
q(x)]\bigr) $$
or, equivalently,
$$P={\rm const.}{\rm exp}\Bigl(k^{-1}\int_{\Omega}dx\bigl[s(q(x))-
{\overline {\theta}}.\bigl(q(x)-{\overline q}\bigr)\bigr]\Bigr).\eqno(2.2.1)$$
Since the function $s$ is concave, the integrand here is maximised when 
$q(x)={\overline q}$ and for small values of the fluctuating field  $\bigl(q(x)-
{\overline q}\bigr)$ is quadratic in this variable. Hence since, as pointed out in 
Section 1, $k$ is a microscopic quantity, extremely small on the macroscopic scale, it 
may be seen from the form of the exponent in Eq. (2.2.1) that $k^{1/2}$ is the natural 
normalisation factor for that field, since that exponent becomes $O(1)$ when $q(x)-
{\overline q}=O(k^{1/2})$. Accordingly, we represent the equilibrium fluctuations 
by the random field ${\xi}$, defined by the formula
$$q(x)={\overline q}+k^{1/2}{\xi}(x).\eqno(2.2.2)$$
In accordance with general desiderata for field theories [SW], we assume that ${\xi}$ 
is a ${\cal D}^{\prime}({\Omega})^{n}$-class distribution, in the sense of L. 
Schwartz [Sc], and we represent its statistical properties by its characteristic function
$${\mu}_{\rm eq}(f)=E_{\rm eq}\bigl({\rm exp}[i{\xi}(f)]\bigr) \ {\forall} \ f{\in}
{\cal D}({\Omega})^{n},\eqno(2.2.3))$$   
where ${\cal D}({\Omega})^{n}$, the dual of ${\cal D}^{\prime}({\Omega})^{n}$,  
is the space of smooth, ${\bf R}^{n}$ valued functions on ${\Omega}$ with support 
in that region, $E_{\rm eq}$ is the expectation functional corresponding to the 
probability distribution $P$ and ${\xi}(f)$ is the smeared field obtained by integrating 
${\xi}$ against the test function $f \ \bigl(=(f_{1},. \ .,f_{n})\bigr)$. We shall 
formulate ${\mu}_{\rm eq}(f)$ in the limit $k{\rightarrow}0$, which we term the 
Boltzmann limit. To this end, we note that, by Eq. (2.2.2), the exponent in Eq. (2.2.1) 
reduces to $\int_{\Omega}dxf(x).[s^{{\prime}{\prime}}({\overline q})f(x)]/2$ in this 
limit. Hence, by Eq. (2..2.3), ${\mu}_{\rm eq}$ is given by the following formula in 
this limit.
$${\mu}_{\rm eq}(f)= {{\int}D{\xi}{\rm exp}
\bigl(\int_{\Omega}dx[{\xi}(x).s^{{\prime}{\prime}}
({\overline q}){\xi}(x)/2+i{\xi}(x).f(x)]\bigr)\over {\int}D{\xi}{\rm exp}
\bigl(\int_{\Omega}dx[{\xi}(x).s^{{\prime}{\prime}}
({\overline q}){\xi}(x)]/2\bigr)},\eqno(2.2.4)$$
where ${\int}D{\xi}(x)$ denotes functional integration over the random field ${\xi}$. 
To provide a proper mathematical definition of this, we proceed as follows. Resolve 
${\Omega}$ into a set of cells, ${\Delta}_{J}$, of equal volume, denote the values of 
${\xi}$ and $f$ at the centroid of ${\Delta}_{J}$ by ${\xi}_{J}$ and $f_{J}$, 
respectively, and re-express Eq. (2.2.4) as
$${\mu}_{\rm eq}(f)={\rm lim}_{\Delta}{\Pi}_{J}
\Bigl[{\int_{R^{n}}d{\xi}_{J}
{\rm exp}\bigl({\xi}_{J}.s^{{\prime}{\prime}}({\overline q}){\xi}_{J}/2 \ 
+i{\xi}_{J}f_{J}\bigr)\over \int_{R^{n}}d{\xi}_{J}
{\rm exp}\bigl({\xi}_{J}.s^{{\prime}{\prime}}
({\overline q}){\xi}_{J}/2\bigr)}\Bigr],\eqno(2.2.5)$$
where ${\rm lim}_{\Delta}$ signifies the limit in which the cells shrink to points. It 
then follows from Eqs. (2.1.12) and  (2.2.5) that 
$${\mu}_{\rm eq}(f)={\rm exp}\bigl(-{1\over 2}(f,{\pi}^{{\prime}{\prime}}
({\overline {\theta}})f)\bigr) \ {\forall} \ f{\in}
{\cal D}({\Omega})^{n}.\eqno(2.2.6)$$
In view of Eq. (2.2.3), this signifies that ${\xi}$ is a Gaussian random field with zero 
mean and two point fumction 
$$E_{eq}\bigl({\xi}_{j}(x){\xi}_{k}(x^{\prime})\bigr)=
{{\partial}^{2}{\pi}({\overline {\theta}})\over 
{\partial}{\overline {\theta}_{j}}{\partial}{\overline {\theta}}_{k}}
{\delta}(x-x^{\prime}) \ {\forall} \ x,x^{\prime}{\in}{\Omega}, \ j,k=1,2. \ 
,n.\eqno(2.2.7)$$
\vskip 0.2cm
In order to obtain the local properties of the random field ${\xi}$ in a global 
equilibrium state, we consider the action of ${\mu}_{\rm eq}$ on test functions that 
are strongly concentrated around an arbitrary point $x_{0}$ of ${\Omega}$. To this 
end we introduce the transformation $f{\rightarrow}f_{x_{0},{\epsilon}}$ of ${\cal 
D}({\Omega})^{n}$, as defined by the formula
$$f_{x_{0},{\epsilon}}(x)={\epsilon}^{-d/2}f\bigl({\epsilon}^{-1}(x-x_{0})\bigr),
\eqno(2.2.8)$$
where ${\epsilon}$ is a \lq small\rq \ positive number. It follows immediately from 
this definition and Eq. (2.2.6) that ${\mu}_{\rm eq}$ is invariant under this 
transformation and therefore the local (punctual !) properties of this characteristic 
function are given by the equation
$${\rm lim}_{{\epsilon}{\downarrow}0}{\mu}_{\rm eq}(f_{x_{0},{\epsilon}})=
{\rm exp}\bigl(-{1\over 2}(f,{\pi}^{{\prime}{\prime}}
({\overline {\theta}})f)\bigr) \ {\forall} \ f{\in}{\cal D}({\Omega})^{n}, \ 
x_{0}{\in}{\Omega}.\eqno(2.2.9)$$ 
\vskip 0.5cm 
\centerline {\bf 2.3. The Quantum Mechanical Picture} 
\vskip 0.3cm
For its microscopic description, we employ the scaling for which the model occupies 
the spatial region ${\Omega}_{L}=L{\Omega}$ and we indicate its $L$ dependence 
by a subscript $L$ to ${\Sigma}$.  In a standard way [VN], we represent its pure 
states and observables by the normalised vectors and the self-adjoint operators, 
respectively, in a separable Hilbert space, ${\cal H}_{L}$, canonically attached to the 
region ${\Omega}_{L}$.  In particular, we denote the Hamiltonian of 
${\Sigma}_{L}$ by $H_{L}$. 
\vskip 0.2cm 
We formulate the statistical thermodynamics of ${\Sigma}_{L}$ in terms of a set of 
extensive, conserved observables ${\hat Q}_{L}=({\hat Q}_{1,L}=H_{L}, \ {\hat 
Q}_{2,L},. \ ,{\hat Q}_{n,L})$. These are designed to be the quantum counterparts of 
the classical variables $Q$, and they intercommute, up to possible surface corrections. 
\vskip 0.2cm
The Von Neumann entropy density of a state represented by the density matrix 
${\rho}_{L}$ is
$${\hat s}_{L}({\rho})=
-L^{-d}k{\rm Tr}\bigl({\rho}_{L}{\rm log}{\rho}_{L}\bigr)$$ 
and the reduced pressure is given by the quantum mechanical version of Eq. (2.1.6), 
namely
$${\pi}_{L}({\theta})=
{\rm sup}_{{\rho}_{L}}\bigl({\hat s}_{L}({\rho}_{L})-{\theta}.{\rm Tr}
\bigl({\rho}_{L}.{\hat Q}_{L})\bigr)=$$
$$-{\rm inf}_{{\rho}_{L}}L^{-d}{\rm Tr}\bigl(k{\rho}_{L}{\rm log}{\rho}_{L}+
{\rho}_{L}.{\hat Q}_{L}\bigr).$$
Hence, as the infimum is attained when ${\rho}_{L}=
{\rm exp}(-k^{-1}{\theta}.{\hat Q}_{L})/{\rm Tr}({\rm idem})$, 
$${\pi}_{L}({\theta})=L^{-d}{\rm log}{\rm Tr}
\bigl({\rm exp}(-k^{-1}{\theta}.{\hat Q}_{L})\bigr)$$
Therefore, in the thermodynamic limit,  the reduced pressure is given by the standard 
formula (cf. [Ru], [Ro1])
$${\pi}({\theta})={\rm lim}_{L\to\infty}{\rm ln}{\rm Tr}
\bigl({\rm exp}(-k^{-1}{\theta}.{\hat Q}_{L})\bigr).\eqno(2.3.1)$$
This formula provides the quantum mechanical infrastructure of the classical 
thermodynamical picture of Section 2.1.
 \vskip 0.2cm
We also formulate the microscopic structure of the equilibrium states of 
${\Sigma}_{L}$ in the thermodynamic limit, wherein it is represented as an infinite 
system, ${\Sigma}_{\infty}$, which occupies the space $X$. We construct the model 
of ${\Sigma}_{\infty}$ in the following standard operator algebraic terms [Ru, Em, 
Se1]. 
\vskip 0.3cm\noindent
{\bf The Operator Algebraic Picture.} We define ${\cal L}$ to be the set of bounded 
open connected regions, ${\lbrace}{\Lambda}{\rbrace}$, of $X$ and we denote the 
volume of  ${\Lambda}$ by ${\vert}{\Lambda}{\vert}$. We represent the 
observables of ${\Sigma}_{\infty}$ that are located in the region ${\Lambda}\ 
({\in}{\cal L})$ by the self-adjoint elements of a $W^{\star}$-algebra, ${\cal 
A}({\Lambda})$, operating in a separable Hilbert space ${\cal H}({\Lambda})$; and 
we assume that ${\cal A}({\Lambda})$ and ${\cal H}({\Lambda})$ satisfy the 
canonical demands of isotony and local commutativity, as given by the formulae
$${\cal H}({\Lambda}_{1}){\subset}{\cal H}({\Lambda}_{2}) \ {\rm and} \ 
{\cal A}({\Lambda}_{1}){\subset}{\cal A}({\Lambda}_{2}) \  {\rm if} \ 
{\Lambda}_{1}{\subset}{\Lambda}_{2}$$
and 
$${\cal A}({\Lambda}_{1}){\subset}{\cal A}({\Lambda}_{2})^{\prime} \ {\rm if} \ 
{\Lambda}_{1}{\cap}{\Lambda}_{2}={\emptyset},$$
respectively, ${\cal A}({\Lambda})^{\prime}$ being the commutant of 
${\cal A}({\Lambda})$. We then define ${\cal A}_{\cal L}$ to be the normed 
$^{\star}$-algebra ${\bigcup}_{{\Lambda}{\in}{\cal L}}{\cal A}({\Lambda})$ and 
${\cal A}$ to be $C^{\star}$-algebra given by its norm completion. These are termed 
the algebras of local and quasi-local bounded observables of ${\Sigma}_{\infty}$, 
respectively. The unbounded local observables of this system are represented by the 
unbounded self-adjoint affiliates of ${\cal A}_{\cal L}$ [Se4].
\vskip 0.2cm
We assume that, for each ${\Lambda}{\in}{\cal L}$, the model has a set of extensive  
observables ${\hat Q}({\Lambda}) \ \bigl(=({\hat Q}_{1}({\Lambda}), \ 
{\hat Q}_{2}({\Lambda}), \ .,{\hat Q}_{n}({\Lambda}))\bigr)$ affiliated to ${\cal 
A}({\Lambda})$, which intercommute, up to possible surface corrections, and are the 
natural counterparts, for the region ${\Lambda}$, of the observables ${\hat Q}_{L}$ 
of ${\Sigma}_{L}$. In particular, ${\hat Q}_{1}({\Lambda})$ is the Hamiltonian of 
the system, ${\Sigma}({\Lambda})$, of particles of the given species confined to 
${\Lambda}$. 
\vskip 0.2cm
We assume that the model is equipped with a representation ${\sigma}$ of the space 
translation group $X$ in ${\rm Aut}({\cal A})$ and that ${\cal A}$  and ${\hat Q}$ 
transform covariantly under these translations, i.e. that 
$${\sigma}(x){\cal A}({\Lambda})={\cal A}({{\Lambda}+x}) \ {\rm and} \ 
{\sigma}(x){\hat Q}({\Lambda})={\hat Q}({\Lambda}+x) \ 
{\forall} \ x{\in}X, \ {\Lambda}{\in}{\cal L}.\eqno(2.3.2)$$ 
\vskip 0.2cm
We take the states of ${\Sigma}_{\infty}$ to comprise the set, ${\cal S}$, of positive, 
normalised, linear functionals, ${\phi}$, on ${\cal A}$, whose restrictions to the local 
algebras ${\cal A}({\Lambda})$ are normal, in that they correspond to density 
matrices, ${\rho}_{\Lambda}^{\phi}$, in the Hilbert spaces ${\cal H}({\Lambda})$ 
according to the formula 
$${\phi}(A)={\rm Tr}({\rho}_{\Lambda}^{\phi}A) \ {\forall} \  
A{\in}{\cal A}({\Lambda}), \ {\Lambda}{\in}{\cal L}.$$ 
This latter condition of local normality is imposed in order to ensure that there is zero 
probability that a bounded spatial region can contain an infinite number of particles 
[DDR]. We represent the action of space translations on ${\cal S}$ by the dual of 
${\sigma}(X)$, we denote by ${\cal S}_{X}$ the set of translationally invariant 
states of the system and by ${\cal S}_{X}^{(0)}$ its subset whose elements, 
${\phi}$, admit well-defined expectation values of the global density of ${\hat Q}$ in 
the form
$${\hat q}({\phi})={\rm lim}_{{\Lambda}{\uparrow}X}{{\rm Tr}
\bigl({\rho}_{\Lambda}^{\phi}{\hat Q}({\Lambda})\bigr)\over 
{\vert}{\Lambda}{\vert}}.\eqno(2.3.3)$$
The functional ${\hat q}$ is manifestly affine. We assume that ${\hat q}({\phi})$ 
represents the value, for the state ${\phi}$, of the classical variable $q$ of Section 2.1 
and thus that the range of ${\hat q}$ is just that of $q$, namely 
${\cal Q}$.. The global entropy density functional, ${\hat s}$, on ${\cal S}_{X}$  is 
defined by the formula [Ru, Ro1]
$${\hat s}({\phi})={\rm lim}_{{\Lambda}{\uparrow}X}
{-k{\rm Tr}\bigl({\rho}_{\Lambda}^{\phi}{\rm ln}({\rho}_{\Lambda}^{\phi})\bigr)
\over {\vert}{\Lambda}{\vert}},\eqno(2.3.4)$$
and this functional is affine. We assume that its relationship to the classical entropy 
functional $s$ of Section 2.1 is given by the formula
$$s(q)={\rm sup}{\lbrace}{\hat s}({\phi}){\vert}{\hat q}({\phi})=q{\rbrace}.
\eqno(2.3.5)$$
Hence the equilibrium condition (2.1.7) may be expressed in the form
$${\pi}({\theta})={\rm sup}{\lbrace}{\hat s}({\phi})-{\theta}.{\hat q}({\phi})
{\vert}{\hat q}({\phi})=q; \ q{\in}{\cal Q}{\rbrace},$$
i.e. 
$${\pi}({\theta})={\hat s}({\phi})-{\theta}.{\hat q}({\phi}).\eqno(2.3.6)$$
This is the GTS condition mentioned in Section 1. We denote by ${\cal G}_{\theta}$ 
the set of GTS states corresponding to the value ${\theta}$ of the control variable.
\vskip 0.3cm\noindent
{\bf Comment.} The above derivation of this GTS condition was based on the 
classical thermodynamic formula (2.1.7) and the identification of ${\hat q}({\rho})$ 
and ${\hat s}({\phi})$ with $q$ and $s(q)$, respectively. The same formula has been 
obtained on purely quantum statistical grounds [Ro1] for lattice systems and for 
continuous systems of particles with hard cores, with ${\hat q}$ comprising the 
densities of energy and particle number.
\vskip 0.2cm
Since both ${\hat s}$ and ${\hat q}$ are affine, it follows from Eq. (2.3.6) that ${\cal 
G}_{\theta}$ is convex. We define ${\cal E}({\cal G}_{\theta})$ to be the set of its 
extremals and assume that these are the pure thermodynamical phases corresponding 
to the value ${\theta}$ of the control variable.. This assumption is supported by the 
observation [Ru] that the extremals are just the GTS states for which the global 
density of ${\hat Q}$ is sharply defined, i.e. dispersionless. Further, by the argument 
of Section 2.1 leading from the convexity of ${\pi}$ to the identification of $-q$ with 
the tangents ${\cal T}_{\theta}$, we see from Eq. (2.3.6) that 
${\cal G}_{\theta}={\lbrace}{\phi}{\in}{\cal S}_{X}^{(0)}{\vert}
{\hat q}({\phi}){\in}-{\cal T}_{\theta}{\rbrace}$. 
\vskip 0.3cm\noindent
{\bf Thermodynamical Completeness Conditions.} We take these conditions, as 
applied to the thermodynamical functional ${\hat q}\bigl(=({\hat q}_{1},. \ .,{\hat 
q}_{n})\bigr)$, and thus to the classical variables $q=(q_{1},. \ .,q_{n})$, to be that
\vskip 0.2cm\noindent
(i) ${\hat q}$ separates the extremals of ${\cal G}_{\theta}$, i.e. if ${\phi}_{1}$ and 
${\phi}_{2}$ are different elements of 
${\cal E}({\cal G}_{\theta})$, then ${\hat q}({\phi}_{1}){\neq}
{\hat q}({\phi}_{2})$; and
\vskip 0.2cm\noindent
(ii) no proper subset of $({\hat q}_{1},. \ .,{\hat q}_{n})$ separates all the extremals 
of ${\cal G}_{\theta}$.
\vskip 0.2cm\noindent
These conditions signify that, for $q{\in}{\cal Q}$, there is a unique extremal GTS 
state for which the value of ${\hat q}$ is $q$. We denote this state by ${\psi}_{q}$. 
In the pure phase regime, where $q=-{\pi}^{\prime}({\theta})$, we denote it, 
equivalently, by ${\omega}_{\theta}$. 
\vskip 0.3cm\noindent
{\bf Dynamics and the KMS Conditions.} Since, by Eqs. (2.1.1) and (2.1.5), the sum 
of the thermal and mechanical energies, $TdS$ and $-pdV$, required to produce an 
increment $dQ$ in the classical variable $Q$ is 
$dE+{\sum}_{j=2}^{n}f_{j}dQ_{j}={\theta}_{1}^{-1}{\theta}.dQ$, we assume 
that, correspondingly, the effective Hamiltonian for the finite version, 
${\Sigma}({\Lambda})$, of ${\Sigma}$ is
$$H_{\theta}^{\Lambda}={\theta}_{1}^{-1}{\theta}.{\hat 
Q}({\Lambda}).\eqno(2.3.7)$$
Hence the dynamics of ${\Sigma}({\Lambda})$ is governed by the one parameter 
group of automorphisms 
${\lbrace}{\alpha}_{\theta}^{\Lambda}({\tau}){\vert}{\tau}{\in}{\bf R}{\rbrace}$ 
of ${\cal A}({\Lambda})$ given by the formula 
$${\alpha}_{\theta}^{\Lambda}({\tau})A=
{\rm exp}(iH_{\theta}^{\Lambda}{\tau}/{\hbar})
A{\rm exp}(-iH_{\theta}^{\Lambda}{\tau}/{\hbar}).\eqno(2.3.8)$$
Since these are inner automorphisms they extend to the whole algebra ${\cal A}$.
\vskip 0.2cm
We now recall that although, in the case of lattice systems, the dynamics of 
${\Sigma}_{\infty}$ may be formulated in terms of automorphisms of ${\cal A}$ 
that are norm limits of ${\alpha}_{\theta}^{\Lambda}$ [St, Ro2], the same is not 
generally true for continuous systems [DS, Ra]. For these, it is only certain privileged 
representations of ${\cal A}$ that support a dynamics that must be based on a weaker 
limiting form of that of ${\Sigma}({\Lambda})$ as ${\Lambda}$ increases to $X$ 
[Se5]. Accordingly, we employ the following scheme to formulate the dynamics of 
${\Sigma}_{\infty}$ in such representations. 
\vskip 0.2cm
We denote by $({\cal H}_{\phi},R_{\phi},{\Psi}_{\phi})$ the GNS triple of a state 
${\phi}$. Then its normal folium ${\cal F}({\phi})$, which comprises the states 
corresponding to the density matrices in ${\cal H}_{\phi}$, is the predual of the 
$W^{\star}$-algebra $R_{\phi}({\cal A})^{{\prime}{\prime}}$. We term this folium 
{\it dynamically amenable} relative to the control parameter ${\theta}$ if it supports a 
dynamics given by a one parameter group of affine transformations 
${\lbrace}{\hat {\alpha}}_{{\star}{\theta}}(t){\vert}t{\in}{\bf R}{\rbrace}$ of 
${\cal F}({\phi})$  given by the $w^{\star}$ limit, as ${\Lambda}$ increases to $X$, 
of the predual of the automorphisms ${\alpha}_{\theta}^{\Lambda}$. Thus
$${\langle}{\hat {\alpha}}_{{\star}{\theta}}(t){\phi}^{\prime};A{\rangle}=
{\rm lim}_{{\Lambda}{\uparrow}X}{\langle}{\phi}^{\prime};
{\alpha}_{\theta}^{\Lambda}(t)A{\rangle} \ {\forall} \ {\phi}^{\prime}{\in}
{\cal F}({\phi}), \ A{\in}{\cal A}, \ t{\in}{\bf R}.\eqno(2.3.9)$$ 
The dual of the transformations ${\hat {\alpha}}_{{\star}{\theta}}$ is then a one 
parameter group ${\lbrace}{\hat {\alpha}}_{\theta}(t){\vert}t{\in}{\bf R}{\rbrace}$
of automorphisms of $R_{\phi}({\cal A})^{{\prime}{\prime}}$ [Se5]. We denote by 
${\hat {\phi}}$ the canonical extension of ${\phi}$ to this algebra. 
\vskip 0.2cm
The KMS conditions on the state ${\phi}$ are that it is dynamically amenable and 
that 
$${\hat {\phi}}\bigl([{\hat{\alpha}}_{\theta}({\tau}){\hat A}]{\hat B}\bigr)=
{\hat {\phi}}\bigl({\hat B}{\hat {\alpha}}_{\theta}({\tau}+i{\hbar}{\beta})
{\hat A}\bigr) \ {\forall} \ {\hat A},{\hat B}{\in}R_{\phi}({\cal 
A})^{{\prime}{\prime}}, \ t{\in}{\bf R},\eqno(2.3.10)$$
where ${\beta}=(kT)^{-1}={\theta}_{1}/k$. 
\vskip 0.2cm
As first proposed  by [HHW], these conditions are designed to characterise the 
equilibrium states of the system, and this proposal is supported by treatments of their 
dynamic [HKT-P] and thermodynamic [PW, Se6] stability. In Section 4 we shall 
assume that the GTS states satisfy these conditions, even though the available proofs 
are confined to lattice systems and a certain class of continuous ones (cf. the 
discussion in [Se1, P.122]).
\vskip 0.5cm 
\centerline {\bf  3. Nonequilibrium Thermodynamics and Local Equilibrium.} 
\vskip 0.3cm
In the previous Section, we saw that, at each level of macroscopicality, the 
equilibrium states are labelled by the value of the classical thermodynamical variable 
$q$. For a  nonequilibrium state, we now take our LTE condition to be that, at each 
level, the local form of the state is the same as for equilibrium, but with that variable 
now taking the space-time dependent form $q(x,t)$ governed by the macroscopic 
evolution of the system. Our objective is to pass from the descriptions of global to 
local equilibrium, at three levels of macroscopicality, in nonequilibrium situations. 
 \vskip 0.3cm
\centerline {\bf 3.1. The Macroscopic Picture.}
\vskip 0.3cm
The phenomenological law  (1.1) is generally based on the following assumptions 
[DM, LL].
\vskip 0.2cm\noindent  
(i) The density $q$ of $Q$ is a function of position $x$ and time $t$ that satisfies a 
local conservation law,
$${{\partial}\over {\partial}t}q(x,t)+{\nabla}.j(x,t)=0,\eqno(3.1.1)$$
where $j=(j_{1},. \ .,j_{n})$ is the associated current density. 
\vskip 0.2cm\noindent
(ii) The model ${\Sigma}$ supports a thermodynamics in which the local entropy 
density is the {\it equilibrium} entropy function, $s$, of  $q(x,t)$. {\it This is the local 
equilibrium assumption.} It serves to define the local control variable, ${\theta}(x,t)$, 
conjugate to $q(x,t)$ by the canonical counterpart of Eq. (2.1.4), namely
$${\theta}(x,t)=s^{\prime}\bigl(q(x,t)\bigr).\eqno(3.1.2)$$
Moreover, the total, time-dependent entropy is 
$$S(t)=\int_{\Omega}dxs\bigl(q(x,t)\bigr).\eqno(3.1.3)$$
\vskip 0.2cm\noindent
(iii) The current $j(x,t)$ is determined by the control variable ${\theta}(x,t)$ 
according to  a constitutive equation of the form
$$j(x,t)={\cal F}({\theta};x,t),\eqno(3.1.4)$$
which, together with Eqs. (3.1.1) and (3.1.2), leads to a positive entropy production 
rate, in accordance with the second law. Moreover, by Eq. (3.1.2), this formula for the 
current may be expressed as 
$$j(x,t)={\cal J}(q;x,t):={\cal F}(s^{\prime}(q);x,t),\eqno(3.1.5)$$ 
and consequently, by Eqs. (3.1.1) and (3.1.5), $q$ evolves autonomously according to 
the  equation 
$${{\partial}\over {\partial}t}q(x,t)+{\nabla}.{\cal J}(q;x,t)=0,\eqno(3.1.6)$$
which is equivalent to Eq. (1.1), with ${\Phi}(q;x,t)=-{\nabla}.{\cal J}(q;x,t)$. 
\vskip 0.2cm
We represent the local thermodynamic state of ${\Sigma}$ by $q(x,t)$, or 
equivalently, in the pure phase regime, by ${\theta}(x,t)$. In fact we shall henceforth 
restrict our considerations to this regime since, as observed in Section 2.3, the 
underlying quantum statistics admits sharp definition of $q$ in pure phases only
\vskip 0.2cm 
We remark that Eqs. (2.1.5) and (3.1.2) suggest that ${\theta}_{1}(x,t)$ is just the 
reciprocal of a local temperature $T(x,t)$: we leave until Section 4  a quantum 
statistical justification of this interpretation that is related to the zeroth law. 
\vskip 0.2cm
To summarise, under the above assumptions (i)-(iii), the macroscopic dynamics of the 
model is given by an autnomous law of the form represented by Eq. (1.1), with 
boundary conditions determined by the reservoirs ${\cal R}$. Specifically, we assume 
that the value of ${\theta}$ at a point of the boundary of ${\Omega}$ is equal to that 
of the corresponding control variable of the reservoir that is in contact with 
${\Sigma}$ there. In particular, we take the equilibrium states of the continuum 
model to be the solutions of Eq. (1.1) for which both $q$ and the imposed boundary 
values of ${\theta}$ are stationary and spatially uniform. 
\vskip 0.5cm 
\centerline {\bf 3.2. Hydrodynamical Fluctuations: the Mesoscopic Picture}
\vskip 0.3cm
We represent the hydrodynamical fluctuations about the deterministic flow $q(x,t)$ 
by a random field ${\xi}_{t}(x)$, normalised, as in the equilibrium situation, by the 
factor $k^{1/2}$. Thus, the local density of $Q$ becomes 
$q(x,t)+k^{1/2}{\xi}_{t}(x)$, which is the natural generalisation of the formula for 
equilibrium fluctuations given by Eq. (2.2.2). As in Section 2.2, we take the random 
field ${\xi}_{t}$ to be a ${\cal D}^{\prime}({\Omega})^{n}$ class distribution. Its 
statistical properties are represented by its characteristic function
$${\mu}(f;t)=E\bigl({\rm exp}(i{\xi}_{t}(f))\bigr) \ {\forall} \ f{\in}
{\cal D}({\Omega})^{n},\eqno(3.2.1)$$
where $E$ is its expectation functional and ${\xi}_{t}(f)$ is the smeared field 
obtained by integrating ${\xi}_{t}$ against $f$. We now assume that, for each time 
$t$, the local properties of ${\mu}$ are just those of ${\mu}_{\rm eq}$, as given by 
Eq. (2.2.9), but with ${\overline {\theta}}$ replaced by ${\theta}(x,t)$. Thus our local 
equilibrium condition for the fluctuating field is that 
$${\rm lim}_{{\epsilon}{\downarrow}0}{\mu}(f_{x,{\epsilon}};t)=
{\mu}_{{\theta}(x,t)}(f),\eqno(3.2.2)$$
where $f_{x,{\epsilon}}$ is defined by Eq. (2.2.8) and
$${\mu}_{{\theta}(x,t)}(f)={\rm exp}\bigl[-{1\over 2}
\bigl(f,{\pi}^{{\prime}{\prime}}\bigl({\theta}(x,t)\bigr)f\bigr)\bigr] \ {\forall} \ 
f{\in}{\cal D}({\Omega})^{n}, \ x{\in}{\Omega}.\eqno(3.2.3)$$ 
We represent the local state of the model at the mesoscopic level by this characteristic 
function 
\vskip 0.2cm
To make the LTE condition perhaps more transparent, we note that, by Eq. (2.2.8), 
$${\xi}_{t}(f_{x,{\epsilon}})={\xi}_{t,x,{\epsilon}}(f),\eqno(3.2.4)$$
where
$${\xi}_{t,x,{\epsilon}}(x^{\prime})=
{\epsilon}^{d/2}{\xi}_{t}(x+{\epsilon}x^{\prime}).\eqno(3.2.5)$$
Hence, by Eqs. (3.2.1), (3.23) and (3.2.5), the formula (3.2.2) may be expressed in the 
form
$${\rm lim}_{{\epsilon}{\downarrow}0}
E\bigl[{\rm exp}\bigl(i{\xi}_{t,x,{\epsilon}}(f)\bigr)\bigr]=
{\rm exp}\bigl({1\over 2}(f,{\pi}^{{\prime}{\prime}}
({\theta}(x,t))f)\bigr) \ {\forall} \ f{\in}{\cal D}({\Omega})^{n}, \ x{\in}{\Omega}, \ 
t{\in}{\bf R}.\eqno(3.2.6)$$
This signifies that, in the punctual limit where ${\epsilon}$ tends to zero, 
${\xi}_{t,x,{\epsilon}}$ becomes a Gaussian random field with zero mean and 
covariance ${\pi}^{{\prime}{\prime}}\bigl({\theta}(x,t)\bigr)$ 
\vskip 0.5cm
\centerline  {\bf 3.3. The Quantum Mechanical Picture.}
\vskip 0.3cm
We consider the state of ${\Sigma}_{L}$ in a region that, in the macroscopic scaling, 
is an open neighbourhood, ${\cal N}({\epsilon},x)$, of an arbitrary point $x$ of 
${\Omega}$, where ${\epsilon}$ is a real positive valued parameter and ${\cal 
N}({\epsilon},x)$ shrinks to the point $x$ as ${\epsilon}{\rightarrow}0$. Then the 
region ${\cal N}({\epsilon},x)$ of the macroscopic picture is just the neighbourhood 
${\cal N}_{L}({\epsilon},x):=L{\cal N}({\epsilon},x)$ of $Lx$ for the microscopic 
one. Thus, for small fixed ${\epsilon}$ and sufficiently large $L$, this region is small 
from the macroscopic standpoint yet large from the microscopic one, indeed 
sufficiently large to contain enormous numbers of particles. It therefore corresponds 
to a hydrodynamic point in the limit where first $L{\rightarrow}{\infty}$ and then 
${\epsilon}{\rightarrow}0$.
\vskip 0.2cm
We denote by ${\tilde {\Omega}}_{L}$ and ${\tilde {\cal N}}_{L}({\epsilon},x)$ 
the regions ${\Omega}_{L}$ and ${\cal N}_{L}({\epsilon},x)$, respectively, as 
viewed relative to the point $Lx$ of ${\Omega}_{L}$. Thus ${\tilde 
{\Omega}}_{L}:={\Omega}_{L}-Lx$ and ${\tilde {\cal 
N}}_{L}({\epsilon},x):={\cal N}_{L}({\epsilon},x)-Lx$. It follows  from these 
specifications that this latter region is infinitely distant from the boundary 
${\partial}{\tilde{\Omega}}_{L}$ of ${\tilde {\Omega}}_{L}$  in the limit 
$L{\rightarrow}{\infty}$; and moreover that, in this limit, it covers the whole space 
$X$
\vskip 0.2cm
The algebra of observables for the region ${\tilde {\cal N}}_{L}({\epsilon},x)$ is 
just the union of the local algebras ${\cal A}({\Lambda})$ of its subregions 
${\Lambda}$ and this reduces to ${\cal A}_{\cal L}$ in the limit 
$L{\rightarrow}{\infty}$. Accordingly, we assume that the local equilibrium 
condition for a pure phase in ${\tilde {\cal N}}_{L}({\epsilon},x)$ is that of GTS 
corresponding to the prevailing value of ${\theta}(x,t)$. Specifically, denoting by 
${\phi}_{L,t}$ the state of ${\Sigma}_{L}$ at time $t$ and by 
${\phi}_{L,t,x,{\epsilon}}$ its restriction to the region ${\tilde {\cal 
N}}_{L}({\epsilon},x)$, we take the local equilibrium condition to be that
$${\rm lim}_{{\epsilon}{\rightarrow}0}
{\lim}_{L\to\infty}{\phi}_{L,t,{\epsilon},x}(A)={\omega}_{{\theta}(x,t)}(A) \  
{\forall} \ A{\in}{\cal A}_{L}.\eqno(3.3.1)$$
where ${\omega}_{\theta}$ is the GTS state of ${\Sigma}_{\infty}$ corresponding 
to the value ${\theta}$ of the control variable in a pure phase, as specified in Section 
2.3. Thus, reverting to the macroscopic scaling, we represent the LTE assumption by 
the attachment of a GTS state ${\omega}_{{\theta}(x,t)}$ to the point $x$ of 
${\Omega}$.
\vskip 0.3cm
{\bf Fibre Bundle Picture.} The situation that we have just decribed may be simply 
represented in terms of the fibre bundle
${\cal B}:={\Omega}_{\bf R}{\times}{\cal A}$, where ${\Omega}_{\bf 
R}:={\Omega}{\times}{\bf R}$. (cf. [SS]). There, the state ${\Phi}$ of 
${\cal B}$ corresponding to the space-time profile ${\theta}(x,t)$ of the macroscopic 
field ${\theta}$ is given by the formula
$${\Phi}\bigl((x,t),A\bigr)={\omega}_{{\theta}(x,t)}(A) \ {\forall} \ 
x{\in}{\Omega}, \ t{\in}{\bf R}, \ A{\in}{\cal A}.\eqno(3.3.2)$$
Evidently the fibre bundle description provides a macroscopic-cum-microscopic 
picture of local equilibrium.
\vskip 0.3cm\noindent
{\bf Local Microdynamics and Local KMS Condition.} We assume here that, as 
envisaged in Section 1, the time scales for the macroscopic and microscopic dynamics 
are infinitely separated. Thus the local macroscopic variable $q(x,t)$ is effectively \lq 
frozen\rq\ at a fixed value and the condition for LTE is that the state of the system in a 
local submacroscopic region is the corresponding equilibrium state.  
\vskip 0.2cm
We now supplement the LTE assumption for the region ${\tilde {\cal 
N}}_{L}({\epsilon},x)$ by a formulation of the microscopic dynamics of the model 
there. For this, we assume that the interactions of the model are of short range. Thus, 
taking account of the facts that, in the limit $L{\rightarrow}{\infty}$, the region 
${\tilde {\cal N}}_{L}({\epsilon},x)$ covers the whole space $X$ and that its algebra 
of observables is ${\cal A}_{\cal L}$, we assume that its microscopic dynamics in 
the normal folium of the local equilibrium state ${\omega}_{{\theta}(x,t)}$ takes the 
same form as for that in the GTS state ${\omega}_{\theta}$ of Section 2.3, with 
${\theta}$ replaced by ${\theta}(x,t)$. On this basis, we assume that, in the limit 
where first $L{\rightarrow}{\infty}$ and then ${\epsilon}{\rightarrow}0$, the 
microscopic dynamics in ${\tilde {\cal N}}_{L}({\epsilon},x)$ is given by the one-
parameter grouip 
${\lbrace}{\hat {\alpha}}_{{\theta}(x,t)}({\tau}){\vert}{\tau}{\in}{\bf R}{\rbrace}$ 
of automorphisms of $R_{{\omega}_{{\theta}(x,t)}}({\cal 
A})^{{\prime}{\prime}}$, as defined in Section 2.3, with 
$R_{{\omega}_{{\theta}(x,t)}}$ the GNS representation of ${\cal A}$ for the state 
${\omega}_{{\theta}(x,t)}$.
\vskip 0.2cm
Correspondingly we assume that, on grounds discussed in Section 2.3, the GTS state 
${\omega}_{{\theta}(x,t)}$ satisfies the KMS condition corresponding to the 
prevailing value, ${\theta}(x,t)$,  of the control parameter. This condition is given by 
the canonical analogue of Eq. (2.3.10), namely
$${\hat {\omega}}_{{\theta}(x,t)}\bigl([{\hat {\alpha}}_{{\theta}(x,t)}({\tau})
{\hat A}]{\hat B}\bigr)={\hat {\omega}}_{{\theta}(x,t)}\bigl({\hat B}
[{\hat {\alpha}}_{{\theta}(x,t)}({\tau}+i{\beta}){\hat A}]\bigr) \ {\forall} \ {\hat A}, 
\ {\hat B}{\in}R_{{\omega}_{{\theta}(x,t)}}({\cal A})^{{\prime}{\prime}}, \ {\tau}
{\in}{\bf R}, $$
$$\ (x,t){\in}{\Omega}_{\bf R},\eqno(3.3.3)$$
where ${\beta}=\bigl(kT(x,t))^{-1}={\theta}_{1}(x,t)/k$. 
This is the local KMS condition.
\vskip 0.5cm 
\centerline {\bf  4. Local Zeroth Law.} 
\vskip 0.3cm
Suppose now that ${\Sigma}$ is coupled to a finite system, ${\Gamma}$, the 
interaction being located within a region which, in the microscopic picture, is a 
bounded, $L$-independent neighbourhood, ${\Delta}$, of the point $Lx$ of 
${\Omega}_{L}$. Thus, for $L$ sufficiently large, ${\Delta}$ is contained in the 
neighbourhood ${\tilde {\cal N}}_{L}({\epsilon},x)$ of $Lx$: moreover, in the 
macroscopic picture, it reduces to the point $x$ in the limit where 
${\epsilon}{\rightarrow}0$. 
\vskip 0.2cm
We now assume the local KMS condition (3.3.3), and recall that the coupling of a 
finite system to an infinite one in a KMS state serves to drive the former to 
equilibrium at the temperature of the latter [KFGV]. It follows that the local 
${\Sigma}-{\Gamma}$ interaction will drive ${\Gamma}$ to equilibrium at 
temperature $T(x,t)$. Thus, under the assumption of local equilibrium, we have a 
local version of the zeroth law
\vskip 0.5cm
\centerline {\bf 5. Concluding Remarks}
\vskip 0.3cm
We have provided  mathematical specifications of local thermodynamic equilibrium 
(LTE)  at macroscopic, mesoscopic and microscopic levels within the framework of 
nonrelativistic quantum statistical thermodynamics.  Furthermore, we provide a 
precise condition (in Section 2.3)  for the thermodynamic completeness of $Q$. 
\vskip 0.2cm
The coordination of our LTE conditions for the single phase regime at the three 
different  levels serves to express them all in terms of ${\theta}(x,t)$. Specifically, it 
represents the local state of ${\Sigma}$ by ${\theta}(x,t)$ at the macroscopic level, 
by the characteristic function ${\mu}_{{\theta}(x,t)}$ at the mesoscopic one (cf. Eq. 
(3.2.3)) and the quantum mechanical form ${\omega}_{{\theta}(x,t)}$ at the 
microscopic level. Furthermore, by considering the local coupling of ${\Sigma}$ to a 
finite one, ${\Gamma}$, we showed that the LTE condition implied a local version of 
the zeroth law.  
\vskip 0.2cm
Our derivation of these results was dependent on the assumption of scale invariance 
of the phenomenological dynamics. This permitted the employment of a 
hydrodynamical limit, which carried an infinite separation of the macroscopic and 
microscopic time scales. However, as noted in Section 2.2, the scale invariance 
assumption is not satisfied in the all-important case of Navier-Stokes fluid mechanics, 
though the ratio of the macroscopic to microscopic time scales is still enormously 
large there. In this and similar cases one may employ the methods of [Se3] to obtain 
the results of the present work, up to miniscule corrections of the order of ratios of 
microscopic to corresponding macroscopic quantities.
\vskip 0.2cm
Finally we remark that, while the present work is designed to be a contribution to 
condensed matter physics, a conceptually different treatment of local equilibrium has 
been formulated in [BOR] within the framework of quantum field theory.
\vskip 0.5cm
\centerline {\bf References}
\vskip 0.3cm\noindent 
[BOR] D. Buchholtz, I. Ojima and H. Roos: {\it Ann. Phys.} {\bf 297} (2002), 219
\vskip 0.2cm\noindent
[Ca]  H. B. Callen: {\it Thermodynamics and an Introduction to Thermostatistics}, 
Wiley, New York, 1985
\vskip 0.2cm\noindent
[Da] E. B. Davies: {\it J. Stat. Phys.} {\bf 18} (1978), 161.
\vskip 0.2cm\noindent
[DDR], G.-F. Dell\rq Antonio, S. Doplicher and D. Ruelle: {\it Commun. Math. 
Phys.} {\bf 2} (1966), 223
\vskip 0.2cm\noindent
[DIPP] A. De Masi, N. Ianiro, A. Pellegrinotti and E. Presutti: {\it From Sochstics to 
Hydrodynamics}, Pp. 127-294 of {\it Nonequilibrium Phenomena II}, Ed. J. L. 
Lebowitz and E. W. Montroll, 1984
\vskip 0.2cm\noindent
[DM]  S. R. De Groot and P. Mazur: {\it Nonequilibrium Thermodynamics}, North 
Holland, Amsterdam, 1962
\vskip 0.2cm\noindent
[DS] D. A. Dubin and G. L. Sewell: {\it J. Math. Phys.} {\bf 11} (1970), 2990
\vskip 0.2cm\noindent
[Em] G. G. Emch: {\it Algebraic Methods in Statistical Mechanics and Quantum 
Field Theory}, Wiley, New York, 1972
\vskip 0.2cm\noindent
[GVV] D. Goderis, A. Verbeure and P. Vets: {\it Prob. Th. Rel. Fields} {\bf 82}, 
527, 1989
\vskip 0.2cm\noindent
[HHW] R. Haag, N. M. Hugenholtz and M. Winnink: {\it Commun. Math. Phys.} 
{\bf 5} (1967), 215
\vskip 0.2cm\noindent
[HKT-P] R. Haag, D. Kastler and E. B. Trych-Pohlmeyer: {\it Commun. Math. 
Phys.} {\bf 38} (1974), 173
\vskip 0.2cm\noindent
[KFGV] A. Kossakowski, A. Frigerio, V. Gorini and M. Verri: {\it Commun. Math. 
Phys.} {\bf 57} (1977), 97
\vskip 0.2cm\noindent
[KMP] C. Kipnis, C. Marchioro and E. Presutti: {\it J. Stat. Phys.} {\bf 27} (1982), 
85
\vskip 0.2cm\noindent
[LL] L. D. Landau and E. M. Lifschitz: {\it Fluid Mechanics}, Pergamon, Oxford, 
1984
\vskip 0.2cm\noindent
[NY] B. Nachtergaele and H-T Yau: {\it Commun. Math. Phys.} {\bf 243} (2003), 
485
\vskip 0.2cm\noindent
[PW] W. Pusz and S. L. Woronowicz: {\it Commun. Math. Phys.} {\bf 58} (1978), 
273
\vskip 0.2cm\noindent
[Pr] E. Presutti: {\it Scaling Limits in Statistical Mechanics and Microstructures in 
Continuum Mechanics}, Springer, Berlin, Heidelberg, 2009
\vskip 0.2cm\noindent
[Ra] C. Radin: {\it Commun. Math. Phys.} {\bf 54} (1977), 69
\vskip 0.2cm\noindent
[Ro1] D. W. Robinson: {\it The Thermodynamic Pressure in Quantum Statistical 
Mechanics: Lecture Notes in Physics} Vol. 9, Springer, Berlin, 1971
\vskip 0.2cm\noindent
[Ro2] D. W. Robinson: {\it Commun. Math. Phys.} {\bf 7} (1968), 337
\vskip 0.2cm\noindent
[Ru] D. Ruelle: {\it Statistical Mechanics}, Benjamin, New York, 1969
\vskip 0.2cm\noindent
[Sc] L. Schwartz: {\it Theorie des Distributions}, Hermann, Paris, 1998
\vskip 0.2cm\noindent
[Se1] G. L. Sewell: {\it Quantum Mechanics and its Emergent Macrophysics}, 
Princeton Univ. Press, Princeton, Oxford, 2002
\vskip 0.2cm\noindent
[Se2] G. L. Sewell: {\it Rev. Math. Phys.} {\bf 17} (2005), 977
\vskip 0.2cm\noindent
[Se3] G. L. Sewell: {\it Rep. Math. Phys.} {\bf 70} (2012), 251
\vskip 0.2cm\noindent
[Se4] G. l. Sewell: {\it J. Math Phys.} {\bf 11} (1970), 1968
\vskip 0.2cm\noindent 
[Se5] G. L. Sewell: {\it Lett. Math. Phys.} {\bf 6} (1982), 209
\vskip 0.2cm\noindent
\vskip 0.2cm\noindent
[Se6] G. L. Sewell: {\it Phys. Rep.} {\bf 57} (1980), 307
\vskip 0.2cm\noindent
[SS] R. N. Sen and G. L. Sewell: {\it J. Math. Phys.} {\bf 43} (2002), 1323  
\vskip 0.2cm\noindent
[St] R. F. Streater: {\it Commun. Math. Phys.} {\bf 6} (1967), 233 
\vskip 0.2cm\noindent
[SW] R. F. Streater and A. S. Wightman: {\it PCT, Spin and Statistics and All That}, 
Benjamin, New York, 1964
\vskip 0.2cm\noindent
[VN] J. Von Neumann: {\it Mathematical Foundations of Quantum Mechanics}, 
Princeton Univ. Press, Princeton, 1955

\end